\documentclass[12pt]{article}
\usepackage{amssymb,amsfonts,amsmath}
\begin{document}
 \hfill\break
\begin{center}\vspace{2.0cm}{\bf Hamilton-Jacobi Treatment of Constraint Field Systems}\\

\vspace{1.0cm}{Walaa I.\ Eshraim}\\
{\textit{New York University Abu Dhabi, Saadiyat Island, P.O. Box 129188, Abu Dhabi, U.A.E.}}
\\
\vspace{1.3cm}{\bf
 Abstract}\end{center} \vspace{0.4cm} Motivated by the Hamilton–Jacobi approach of fields with constraints, we analyse the classical structure of three different constrained field systems: (i) the scalar
field coupled to two flavours of fermions through Yukawa
couplings (ii) the scalar field coupled minimally to the vector potential (iii) the electromagnetic field coupled to a spinor. The equations of motion are obtained as
total differential equations in many variables. The integrability conditions are investigated. The second and third constrained systems are quantized using canonical path integral formulation based on the Hamilton-Jacobi treatment.
 
 \begin{center}{\bf{keywords}}: Hamiltonian-Jacobi formalism, constrained systems, path integral.\\
\end{center}

 \vfill\eject
\section{Introduction}
The most common method for investigating the Hamiltonian treatment of constrained systems was initiated by Dirac [1,2]. The main feature of his method is to consider primary constraints first. All constraints are obtained using consistency conditions. Besides, he showed that the number of degrees of freedom of the dynamical system can be reduced. Hence, the equations of motion of a constrained system are obtained in terms of arbitrary parameters. Moreover, the Dirac approach is widely used for quantizing the
constrained Hamilton systems. The path integral is another
approach used for the quantization of constrained systems of
classical singular theories, which was initiated by Faddeeve [3].
Faddeeve has applied this approach when only first-class
constraints in the canonical gauge are present. Senjanovic [4]
generalized Faddeev's method to second-class constraints. Fradkin
and Vilkovisky [5,6] red-rived both results in a broader context,
where they improved the procedure to the Grassman variables. Gitman
and Tyutin [7] discussed the canonical quantization of singular
theories as well as the Hamiltonian formalism of gauge theories in
an arbitrary gauge.\\
\indent The canonical method (or G\"{u}ler's method) developed Hamilton-Jacobi formulation to investigate constrained systems [8-9]. The starting point of the Hamilton-Jacobi approach [10-14] is the variational principle. The Hamiltonian treatment of constrained systems leads us to the equations of motion as total differential equations in many variables. The equations are integrable if the corresponding system of partial differential equations is a Jacobi system. In Ref. [15] G\"{u}ler has presented a treatment of classical fields as constrained systems. Then Hamilton-Jacobi quantization of finite dimensional system with constraints was investigated in Ref. [16]. The advantages of the Hamilton-Jacobi formalism [17-21] are that there are no differences between first- and second-class constraints and no need for a gauge-fixing term because the gauge variables are separated in the processes of constructing an integrable system of total differential equations. The Hamilton-Jacobi approach treats the constrained system as the same as Dirac's methods but in a simple way. Both methods give the same results as seen in Refs. [22-26]. \\
\indent This work is organized as follows: In Sec.2 Hamilton-Jacobi formulation is presented. In Sec.3 the Hamilton-Jacobi formulation of the scalar field coupled to two flavours of fermions through Yukawa couplings is investigated. In Sec.4 the path integral quantization of the scalar field coupled minimally to the vector potential is obtained In Sec.5 the system as the electromagnetic field coupled to a spinor is quantized using Hamilton-Jacobi formulation. In Sec.4 The conclusions are given.\\

\section{Hamilton-Jacobi Formulation}
\indent One starts from singular Lagrangian  $L\equiv
L(q_{i},{\dot{q}}_{i},t),\> i=1,2,\ldots,n$, with the Hess matrix
\begin{equation}\label{1}
A_{ij}=\frac{\partial^{2}L}{{\partial{\dot{q}}_{i}}\>{\partial{\dot{q}}_{j}}}\>,
\end{equation}
of rank $(n-r)$, $r<n$. The generalized momenta $p_{i}$
corresponding to the generalized coordinates $q_{i}$ are defined
as
\begin{align}
p_{a}&= \frac{\partial{L}}{\partial{\dot{q}}_{a}},\qquad {a =
1,2,\ldots,n-r}, \label{2}\\
p_{\mu}&= \frac{\partial{L}}{\partial{\dot{x}}_{\mu}},\qquad {\mu
= n-r+1,\ldots,n}.\label{3}
\end{align}
where $q_{i}$ are divided into two sets, $q_{a}$ and $x_{\mu}$.
Since the rank of the Hessian matrix is $(n-r)$, one may solve
Eq. (4) for ${\dot{q}}_{a}$ as
\begin{equation}\label{4}
{\dot{q}}_{a}={\dot{q}}_{a}(q_{i},{\dot{x}}_{\mu},p_{a};t).
\end{equation}
Substituting Eq. (4), into Eq. (3), we get
\begin{equation}\label{5}
p_{\mu}=-H_{\mu}(q_{i},{\dot{x}}_{\mu},p_{a};t).
\end{equation}
 The canonical Hamiltonian $H_{0}$ reads
\begin{equation}\label{6}
H_{0}=-L(q_{i},{\dot{x}}_{\nu},{\dot{q}}_{a};t)+
p_{a}{\dot{q}}_{a}-{\dot{x}}_{\mu}H_{\mu},\qquad
{\nu=1,2,\ldots,r}.
\end{equation}
The set of Hamilton-Jacobi Partial Differential Equations is
expressed as
\begin{equation}\label{7}
H'_{\alpha}\bigg(x_{\beta},~q_{\alpha},\frac{\partial S}{\partial
q_{\alpha}} ,\frac{\partial S}{\partial
x_{\beta}}\bigg)=0,\qquad{\alpha,\beta=0,1,\ldots,r},
\end{equation}
where
\begin{equation}\label{8}
H'_{0}= p_{0}+H_{0}~,
\end{equation}
\begin{equation}\label{9}
 H'_{\mu}= p_{\mu}+H_{\mu}~.
\end{equation}
We define $p_{\beta}=\partial S[q_{a};x_{a}]/\partial x_{\beta}$
and $p_{a}=\partial S[q_{a};x_{a}]/\partial q_{a}$ with $x_{0}=t$
and $S$ being the action.\\
\indent Now the total differential equations are given as
\begin{align}
dq_{a}&=\,\frac{\partial H'_{\alpha}}{\partial p_{a}}\,dx_{\alpha},\label{10}\\
dp_{a}&=\frac{\partial H'_{\alpha}}{\partial q_{a}}\,dx_{\alpha},\label{11}\\
dp_{\beta}&=\frac{\partial H'_{\alpha}}{\partial
t_{\beta}}\,dx_{\alpha}, \label{12}
\end{align}
\begin{equation}\label{13}
dz= \bigg(-H_{\alpha}+p_{a}\frac{\partial H'_{\alpha}}{\partial
p_{a}}\bigg) dx_{\alpha},
\end{equation}
where $Z=S(x_{\alpha},q_{a})$. These equations are integrable if
and only if [27]
\begin{equation}\label{14}
dH'_{0}=0,
\end{equation}
\begin{equation}\label{15}
dH'_{\mu}=0,\,\;\quad \quad \mu=1,2,\ldots,r.
\end{equation}
If conditions (14), (15) are not satisfied identically, one
considers them as a new constants and a gain consider their
variations. Thus, repeating this procedure, one may obtain a set
of constraints such that all variations vanish. Simultaneous
solutions of canonical equations with all these constraints
provide the set of canonical phase space coordinates
$(q_{a},p_{a})$ as functions of $t_{a}$; the canonical action
integral is obtained in terms of the canonical coordinates.
$H'_{\alpha}$ can be interpreted as the infinitesimal generator of
canonical transformations given by parameters $t_{\alpha}$,
respectively. In this case the path integral representation can be
written as [28, 29].
\begin{equation}\label{16}
\left<Out\mid S\mid In\right>  =\int\prod_{a=1}^{n-p} dq^{a}
dp^{a} \exp{\left[ i\int_{t_{\alpha}}^{t'_{\alpha}}{\left(
-H_{\alpha}+ p_{a}\frac{\partial H'_{\alpha}}{\partial p_{a}}
\right)} dt_{\alpha}\right]},
\end{equation}
\indent $\,\,\,\quad \quad \quad \quad a=1,,\ldots,n-p,\quad \quad
\alpha=0,n-p+1,\ldots,n$.\\
\indent In fact, this path integral is an integration over the
canonical phase space coordinates $(q^{a},p^{a})$.

%%%%%%%%%%%%%%%%%%%%%%
\section{Hamilton-Jacobi formulation of the scalar
field coupled to two flavours of fermions through Yukawa
couplings}
We consider one loop order the self-energy for the scalar field
$\varphi$ with a mass $m$, coupled to two flavours of fermions
with masses $m{_1}$ and $m{_2}$, coupled through Yukawa couplings
described by the Lagrangian
\begin{multline}\label{17}
L={\frac{1}{2}}(\partial_{\mu}\varphi)^2-{\frac{1}{2}}{m^2}{\varphi^2}-{\frac{1}{6}}\lambda
{\varphi^3}+\sum_{i}{{\overline{\psi}}_{(i)}}(i{\gamma^\mu}\partial_\mu-m_i){\psi_{(i)}}\\
-g\varphi({{\overline{\psi}}_{(1)}}{\psi_{(2)}}+{{\overline{\psi}}_{(2)}}{\psi_{(1)}}),
\qquad{\mu =0,1,2,3},
\end{multline}
where $\lambda$ is parameter and $g$ constant, $\varphi$,
$\psi_{(i)}$, and ${\overline\psi}_{(i)}$ are odd ones. We are
adopting the Minkowski metric
$\eta_{\mu\nu}=diag(+1,-1,-1,-1)$.\\
\indent The Lagrangian function (17) is singular, since the rank
of the Hess matrix
\begin{equation}\label{18}
A_{ij}=\dfrac{\partial^{2}L}{{\partial{\dot{q}}_{i}}\>
{\partial{\dot{q}}_{j}}},
\end{equation}is one.\\
The generalized momenta (2,3)
\begin{equation}\label{19}
p_{\varphi} =\frac{\partial
L}{\partial\dot{\varphi}}={\partial^0}\varphi,
\end{equation}
\begin{equation}\label{20}
p_{(i)} =\frac{\partial L}{\partial{\dot{\psi}}_{(i)}}=
i{{\overline{\psi}}_{(i)}}{\gamma^0}= -H_{(i)},\qquad {i =1,2},
\end{equation}
\begin{equation}\label{21}
\overline{p}_{(i)}=\frac{\partial
L}{\partial{\dot{\overline\psi}}_{(i)}}=0=-\overline{H}_{(i)}.
\end{equation}
Where we must call attention to the necessity of being careful
with the spinor indexes. Considering, as usual $\psi_{(i)}$ as a
column vector and ${\overline\psi}_{(i)}$ as a row vector implies
that $p_{(i)}$ will be a row vector while ${\overline p}_{(i)}$
will be a column vector. \\
 Since the rank of the Hess matrix is one, one may
solve (19) for ${\partial^0}\varphi$ as:
\begin{equation}\label{22}
{\partial^0}\varphi=p_{\varphi}\equiv\omega.
\end{equation}
The usual Hamiltonian $H_0$ is given as:
\begin{equation}\label{23}
H_0 = -L+\omega
p_{\varphi}+{{\partial_0}\psi_{(i)}}\>p_{(i)}\>{\bigg|_{{p_{(i)}=-H_{(i)}}}}+{
{{\partial_0}{\overline{\psi}_{(i)}}}\>{\overline
p}_{(i)}}\>\bigg|_ {{\overline p}_{(i)}= -{\overline H}_{(i)}},
\end{equation}
or
\begin{multline}\label{24}
H_0 =
{\frac{1}{2}}({p^2}_{\varphi}-{\partial_{a}\varphi}{\partial^a}\varphi)+{\frac{1}{2}}{m^2}{\varphi^2}+{\frac{1}{6}}\lambda
{\varphi^3}-{{\overline{\psi}}_{(i)}}(i{\gamma^a}\partial_{a}-m_i){\psi_{(i)}}
\\+g\varphi({{\overline{\psi}}_{(1)}}{\psi_{(2)}}+{{\overline{\psi}}_{(2)}}{\psi_{(1)}}),\qquad {a=1,2,3.}
\end{multline}
Eqs. (20), and (21) lead to the primary constraints.\\

By using Hamilton-Jacobi, the set of (HJPDE) (8) read as
\begin{multline}\label{58}
H'_0 =p_0+H_0=p_0+
{\frac{1}{2}}({p^2}_{\varphi}-{\partial_{a}\varphi}{\partial^a}\varphi)+{\frac{1}{2}}{m^2}{\varphi^2}+{\frac{1}{6}}\lambda
{\varphi^3}-{{\overline{\psi}}_{(i)}}(i{\gamma^a}\partial_{a}-m_i){\psi_{(i)}}\\
+g\varphi({{\overline{\psi}}_{(1)}}{\psi_{(2)}}+{{\overline{\psi}}_{(2)}}{\psi_{(1)}}),
\end{multline}
\begin{equation}\label{59}
H'_{(i)}=p_{(i)}+H_{(i)}=p_{(i)}-i\>{{\overline{\psi}}_{(i)}}\,{\gamma^0}=0,
\end{equation}

\begin{equation}\label{60}
\overline H'_{(i)}=\overline p_{(i)}+\overline H_{(i)}=\overline
p_{(i)}=0.
\end{equation}

Therefore, the total differential equations for the characteristic
(10), (11) and (12) are:
\begin{equation}\label{28}
d\varphi=p_{\varphi}d\tau,
\end{equation}

\begin{equation}\label{29}
d{\psi_{(i)}}=d{\psi_{(i)}},
\end{equation}

\begin{equation}\label{30}
d{{\overline\psi}_{(i)}}=d{{\overline\psi}_{(i)}},
\end{equation}

\begin{equation}\label{31}
d{{p}_{\varphi}}=\bigg[{m^2}\varphi+{\frac{1}{2}}\lambda{\varphi^2}
+g({{\overline{\psi}}_{(1)}}{\psi_{(2)}}+{{\overline{\psi}}_{(2)}}{\psi_{(1)}})\bigg]d\tau,
\end{equation}

\begin{equation}\label{32}
d{{p}_{(1)}}=\bigg[{\overline\psi}_{(1)}(i\overleftarrow{\partial_a}\gamma^a
+m_{1})+g\,\varphi\,{\overline\psi}_{(2)}\bigg]d\tau,
\end{equation}

\begin{equation}\label{33}
d{{p}_{(2)}}=\bigg[{\overline\psi}_{(2)}(i\overleftarrow{\partial_a}\gamma^a
+m_{2})+g\,\varphi\,{\overline\psi}_{(1)}\bigg]d\tau,
\end{equation}

\begin{equation}\label{34}
d{\overline
p_{(1)}}=\bigg[-(i{\gamma^a}{\partial_a}-m_{1}){\psi_{(1)}}+g\varphi{\psi_{(2)}}\bigg]d\tau
-i{\gamma^0}d{\psi_{(1)}},
\end{equation}

\begin{equation}\label{35}
d{\overline
p_{(2)}}=\bigg[-(i{\gamma^a}{\partial_a}-m_{2}){\psi_{(2)}}+g\varphi{\psi_{(1)}}\bigg]d\tau
-i{\gamma^0}d{\psi_{(2)}}.
\end{equation}
\\
The integrability conditions $(dH'_{\alpha}=0)$ imply that the
variation of the constraints $H'_{(i)}$ and $\overline H'_{(i)}$
should be identically zero, that is
\begin{equation}\label{36}
dH'_{(i)}=dp_{(i)}-i\>d{{\overline{\psi}}_{(i)}}\,{\gamma^0}=0,
\end{equation}

\begin{equation}\label{37}
d{\overline H'_{(i)}}=d\overline p_{(i)}=0.
\end{equation}
\indent The following equations of motion:\\
 From Eq. (28), we obtain
\begin{equation}\label{38}
\dot\varphi= p_{\varphi}.
\end{equation}
 Substituting from Eqs. (32) and (33) into Eq. (36), we get

\begin{equation}\label{39}
i\partial_{0}{\overline\psi}_{(1)}\gamma^{0}-{\overline\psi}_{(1)}(i\overleftarrow{\partial_a}\gamma^a
+m_{1})-g\varphi\,{\overline\psi}_{(2)}=0,
\end{equation}

\begin{equation}\label{40}
i\partial_{0}{\overline\psi}_{(2)}\gamma^{0}-{\overline\psi}_{(2)}(i\overleftarrow{\partial_a}\gamma^a
+m_{2})-g\varphi\,{\overline\psi}_{(1)}=0.
\end{equation}
Substituting from Eqs. (34) and (35) into Eq. (37), we have

\begin{equation}\label{41}
(i{\gamma^\mu}{\partial_\mu}-m_{1}){\psi_{(1)}}-g\,\varphi\>{\psi_{(2)}}=0,
\end{equation}

\begin{equation}\label{42}
(i{\gamma^\mu}{\partial_\mu}-m_{2}){\psi_{(2)}}-g\,\varphi\>{\psi_{(1)}}=0.
\end{equation}
One notes that the integrability conditions are not identically
zero, they are added to the set of equations of motion.\\
From Eqs.(31-33), we get the following equations of motion:
\begin{equation}\label{43}
{\dot p}_{\varphi}={m^2}\varphi+{\frac{1}{2}}\lambda{\varphi^2}
+g({{\overline{\psi}}_{(1)}}{\psi_{(2)}}+{{\overline{\psi}}_{(2)}}{\psi_{(1)}}),
\end{equation}

\begin{equation}\label{44}
{\dot
p}_{(1)}={\overline\psi}_{(1)}(i\overleftarrow{\partial_a}\gamma^a
+m_{1})+g\,\varphi\,{\overline\psi}_{(2)},
\end{equation}

\begin{equation}\label{45}
{\dot
p}_{(2)}={\overline\psi}_{(2)}(i\overleftarrow{\partial_a}\gamma^a
+m_{2})+g\,\varphi\,{\overline\psi}_{(1)}.
\end{equation}
\\
Substituting from Eqs. (41) and (42) into Eqs. (34) and (35), we get
\\
\begin{equation}\label{46}
{\dot{\overline p}}_{(i)}=0,\qquad {i =1,2}.
\end{equation}
\\
Differentiate Eq. (38) with respect to time, and making use of
Eq. (43), we get
\begin{equation}\label{47}
\ddot{\varphi}-{m^2}\varphi-{\frac{1}{2}}\lambda{\varphi^2}
-g({{\overline{\psi}}_{(1)}}{\psi_{(2)}}+{{\overline{\psi}}_{(2)}}{\psi_{(1)}})=0.
\end{equation}
%%%%%%%%%%%%%%%%%

\section{Path integral quantization of the scalar field coupled minimally to the vector potential}
\indent Consider the action integral for the scalar field coupled
minimally to the vector potential as
\begin{equation}\label{48}
S=\int d_{4}x \>L,
\end{equation}
where the Lagrangian $L$ is given by
\begin{equation}\label{49}
L=-\frac{1}{4}\>F_{\mu\nu}(x) F^{\mu\nu}(x)+
(D_{\mu}\varphi)^\ast(x)D^{\mu}\varphi(x)-m^2
\varphi^{\ast}(x)\varphi(x),
\end{equation}
where
\begin{equation}\label{50}
F^{\mu\nu}=\partial^{\mu}A^{\nu}-\partial^{\nu}A^{\mu},
\end{equation}
and
\begin{equation}\label{51}
D_{\mu}\varphi(x)=\partial_{\mu}\varphi(x)-ieA_{\mu}(x)\varphi(x).
\end{equation}
The canonical momenta are defined as
\begin{equation}\label{52}
\pi^{i}=\frac{\partial L}{\partial {\dot{A}}_{i}}=-F^{0i},
\end{equation}
\begin{equation}\label{53}
\pi^{0}= \frac{\partial L}{\partial{\dot{A}}_{0}}=0,
\end{equation}
\begin{equation}\label{54}
p_{\varphi}= \frac{\partial L}{\partial
\dot{\varphi}}=(D_{0}\varphi)^{\ast}=\dot{\varphi}^{\ast}+ieA_{0}\varphi^{\ast},
\end{equation}
\begin{equation}\label{55}
p_{\varphi^{\ast}}= \frac{\partial L}{\partial
{\dot{\varphi}}^{\ast}}=(D_{0}\varphi)=\dot{\varphi}-i\,e\,A_{0}\,\varphi,
\end{equation}
\indent From Eqs. (52), (54) and (55), the velocities
${\dot{A}}_{i},{\dot{\varphi}}^{\ast}$ and $\dot{\varphi}$ can be
expressed in terms of momenta $\pi_{i}, p_{\varphi}$ and
$p_{\varphi^{\ast}}$ respectively as
\begin{equation}\label{56}
{\dot{A}}_{i}=-\pi_{i}-\partial_{i}A_{0},
\end{equation}
\begin{equation}\label{57}
{\dot{\varphi}}^{\ast}=p_{\varphi}-ieA_{0}{\varphi}^{\ast},
\end{equation}

\begin{equation}\label{58}
{\dot{\varphi}}=p_{{\varphi}^{\ast}}+ieA_{0}{\varphi}.
\end{equation}
The canonical Hamiltonian $H_{0}$ is obtained as
\begin{multline}\label{59}
\qquad
H_{0}=\frac{1}{4}\>F^{ij}F_{ij}-\frac{1}{2}\>\pi_{i}\pi^{i}+\pi^{i}\,\partial_{i}A_{0}+p_{{\varphi}^{\ast}}p_{\varphi}+ieA_{0}{\varphi}p_{\varphi}
\\\,\,\qquad\qquad-ieA_{0}{\varphi}^{\ast}p_{{\varphi}^{\ast}}-(D_{i}\varphi)^{\ast}(D^{i}\varphi)
+m^{2}{\varphi}^{\ast}\varphi.
\end{multline}\\
Making use of Eqs. (7) and (9), we find for the set of HJPDE
\begin{equation}\label{60}
H'_{0}=\pi_{4}+H_{0},
\end{equation}
\begin{equation}\label{61}
H'=\pi_{0}+H=\pi_{0}=0,
\end{equation}
Therefore, the total differential equations for the characteristic
(10-12) obtained as
\begin{multline}
\quad \quad \quad \quad\qquad\qquad\quad dA^{i}=\frac{\partial
H'_{0}}{\partial\pi_{i}}\>dt+\frac{\partial
H'}{\partial\pi_{i}}\>dA^0,\\
=-(\pi^{i}+\partial_{i}A_{0})\,dt,\quad \quad \quad \quad \quad
\quad \quad \quad \quad \,\,\,\,\,\label{62}
\end{multline}

\begin{equation}\label{63}
dA^{0}=\frac{\partial H'_{0}}{\partial\pi_{0}}\>dt+\frac{\partial
H'}{\partial\pi_{0}}\>dA^0=dA^0,
\end{equation}
\begin{multline}
\quad \quad \quad \quad\qquad\qquad\quad d\varphi=\frac{\partial
H'_{0}}{\partial p_{\varphi}}\>dt+\frac{\partial
H'}{\partial p_{\varphi}}\>dA^0,\\
=(p_{{\varphi}^{\ast}}+ieA_{0}\varphi)\,dt,\quad \quad \quad \quad
\quad \quad \quad \quad \quad \,\,\,\,\,\label{64}
\end{multline}
\begin{multline}
\quad \quad \quad \quad\qquad\qquad\quad
d\varphi^{\ast}=\frac{\partial H'_{0}}{\partial
p_{\varphi^{\ast}}}\>dt+\frac{\partial
H'}{\partial p_{\varphi^{\ast}}}\>dA^0,\\
=(p_{\varphi}-ieA_{0}\varphi^{\ast})\,dt,\quad \quad \quad \quad
\quad \quad \quad \quad \quad \,\,\label{65}
\end{multline}
\begin{multline}
\quad \quad \quad \quad d\pi^{i}=-\frac{\partial H'_{0}}{\partial
A_{i}}\>dt-\frac{\partial H'}{\partial A_{i}}\>dA^0,\\
=[\partial_{l}F^{li}+ie(\varphi^{\ast}\partial^{i}\varphi+\varphi\,\partial_{i}\varphi^{\ast})+2e^{2}A^{i}\varphi\varphi^{\ast}]\,dt,\quad
\quad \quad \,\,\label{66}
\end{multline}
\begin{multline}
\quad \quad \quad \quad \quad \quad\,\,\,\,
d\pi^{0}=-\frac{\partial H'_{0}}{\partial
A_{0}}\>dt-\frac{\partial H'}{\partial A_{0}}\>dA^0,\\
=[\partial_{i}\pi^{i}+ie\varphi^{\ast}p_{{\varphi}^{\ast}}-ie\varphi\,p_{\varphi}]\,dt,\quad
\quad \quad \quad \quad \quad \quad\, \label{67}
\end{multline}
\begin{multline}
\quad \quad \quad \quad \quad \quad\,\,\,\,
dp_{\varphi}=-\frac{\partial H'_{0}}{\partial
\varphi}\>dt-\frac{\partial H'}{\partial \varphi}\>dA^0,\\
=[(\overrightarrow{D}\cdot\overrightarrow{D}\varphi)^{\ast}-m^{2}\varphi^{\ast}-ieA_{0}p_{\varphi}]\,dt,
\quad \quad \quad \quad \,\,\,\,\,\label{68}
\end{multline}
and
\begin{multline}
\quad \quad \quad \quad \quad \quad\,\,\,\,
dp_{\varphi^{\ast}}=-\frac{\partial H'_{0}}{\partial
\varphi^{\ast}}\>dt-\frac{\partial H'}{\partial \varphi^{\ast}}\>dA^0,\\
=[(\overrightarrow{D}\cdot\overrightarrow{D}\varphi)-m^{2}\varphi+ieA_{0}p_{\varphi^{\ast}}]\,dt.
\quad \quad \quad \quad \,\,\,\,\,\label{69}
\end{multline}
\indent The integrability condition $(dH'_{\alpha}=0)$ implies
that the variation of the constraint $H'$ should be identically
zero, that is
\begin{equation}\label{70}
dH'=d\pi_{0}=0,
\end{equation}
which leads to a new constraint
\begin{equation}\label{71}
H''=\partial_{i}\pi^{i}+ie\varphi^{\ast}p_{{\varphi}^{\ast}}-ie\varphi\,p_{\varphi}=0.
\end{equation}
Taking the total differential of $H''$, we have
\begin{equation}\label{72}
dH''=\partial_{i}d\pi^{i}+iep_{{\varphi}^{\ast}}d\varphi^{\ast}+ie\varphi^{\ast}dp_{{\varphi}^{\ast}}-ie\varphi\,dp_{\varphi}-iep_{\varphi}\,d\varphi=0.
\end{equation}
Then the set of equation (62-69) is integrable. Therefore, the
canonical phase space coordinates $(\varphi,p_{\varphi})$ and
$(\varphi^{\ast},p_{\varphi^{\ast}})$ are obtained in terms of
parameters $(t,A^{0})$.\\
\indent Making use of Eq. (13) and (59-61), we obtain the canonical
action integral as
\begin{multline}\label{73}
\quad Z=\int
d^{4}x(-\frac{1}{4}\>F^{ij}F_{ij}-\frac{1}{2}\>\pi_{i}\pi^{i}+p_{\varphi}p_{\varphi^{\ast}}+\overrightarrow{D}\varphi^{\ast}\cdot\overrightarrow{D}\varphi+m^{2}|\varphi|^2),
\end{multline}
where
\begin{equation}\label{74}
\overrightarrow{D}=\overrightarrow{\bigtriangledown}+ie\overrightarrow{A}.
\end{equation}
Now the path integral representation (16) is given by
\begin{multline}\label{75}
\left<out|S|In\right> = \int\prod_{i}\,dA^{i}\,d\pi^{i}\,
d\varphi\,
dp_{\varphi}\,d\varphi^{\ast}\,dp_{\varphi^{\ast}}\>exp\,\bigg[i\bigg\{\int d^{4}x\\
(-\frac{1}{2}\>\pi_{i}\pi^{i}-\frac{1}{4}\>F^{ij}F_{ij}+p_{\varphi}p_{\varphi^{\ast}}
+(D_{i}\varphi)^{\ast}(D_{i}\varphi)-m^{2}\varphi^{\ast}\varphi)\bigg\}\bigg].
\end{multline}

%%%%%%%%%%%%%%%%%%%%%%%%%%%%%%%%%%%%%%%

\section{Path integral quantization of electromagnetic field coupled to a spinor}

\indent We analyse the case of the electromagnetic field
coupled to a spinor, whose Hamiltonian formalism was analysed
[30,31]. We will consider the Lagrangian density written as
\begin{equation}\label{76}
L=-\frac{1}{4}\>F_{\mu\nu}F^{\mu\nu}+
i\overline{\psi}\gamma^{\mu}(\partial_{\mu}+ieA_{\mu})\psi-m\overline{\psi}\psi,
\end{equation}
where $A_{\mu}$ are even variables while $\psi$ and
$\overline{\psi}$ are odd ones. The electromagnetic tensor is
defined as
$F^{\mu\nu}=\partial^{\mu}A^{\nu}-\partial^{\nu}A^{\mu}$ and we
are adopting the Minkowski metric
$\eta_{\mu\nu}=diag(+1,-1,-1,-1)$.\\
The Lagrangian function (76) is singular, since the rank of the
Hess matrix
\begin{equation}\label{77}
A_{ij}=\frac{\partial^{2}L}{{\partial{\dot{q}}_{i}}\>{\partial{\dot{q}}_{j}}}\>,
\end{equation}
is three.\\
\indent The momenta variables conjugated, respectively, to $A_i,
A_0, \psi$ and $\overline\psi$, are
\begin{equation}\label{78}
\pi^{i}=\frac{\partial L}{\partial {\dot{A}}_{i}}=-F^{0i},
\end{equation}
\begin{equation}\label{79}
\pi^{0}= \frac{\partial L}{\partial{\dot{A}}_{0}}=0=-H_1,
\end{equation}
\begin{equation}\label{80}
p_{\psi}= \frac{\partial L}{\partial
\dot{\psi}}=i\>\overline{\psi}\gamma^{0}=-H_{\psi}\>,
\end{equation}
\begin{equation}\label{81}
p_{\overline{\psi}}= \frac{\partial L}{\partial
\dot{\overline{\psi}}}=0=-H_{\overline{\psi}}\>,
\end{equation}
where we must call attention to the necessity of being careful
with the spinor indexes. Considering, as usual, $\psi$ as a column
vector and $\overline\psi$ as a row vector implies that $p_\psi$
will be a row vector while $p_{\overline\psi}$ will be a column
vector.\\
With the aid of relation (78), the Lagrangian density may be
written as
\begin{equation}\label{82}
L=-\frac{1}{2}\>\pi_{i}\pi^i-\frac{1}{4}\>F_{ij}F^{ij}+
i\overline{\psi}\gamma^{\mu}(\partial_{\mu}+ieA_{\mu})\psi-m\overline{\psi}\psi,
\end{equation}
then the canonical Hamiltonian density takes the form
\begin{multline}\label{83}
H_{0}=\pi^i\dot{A}_i+\frac{1}{2}\>\pi_{i}\pi^{i}+\frac{1}{4}\>F^{ij}F_{ij}-i\overline{\psi}(\gamma^{\mu}ieA_{\mu}+
\gamma^{i}\partial_{i})\psi+m\overline{\psi}\psi.
\end{multline}
The velocities $\dot{A}_i$ can be expressed in terms of the
momenta $\pi_{i}$ as
\begin{equation}\label{84}
{\dot{A}}_{i}=-\pi_{i}+\partial_{i}A_{0}.
\end{equation}
Therefore, the Hamiltonian density is
\begin{multline}\label{85}
H_{0}=\frac{1}{4}\>F^{ij}F_{ij}-\frac{1}{2}\>\pi_{i}\pi^{i}+\pi^i\partial_iA_0+
\overline{\psi}\gamma^{\mu}eA_{\mu}\psi-\overline{\psi}(i\gamma^i\partial_i-m)\psi.
\end{multline}
\indent The set of Hamilton-Jacobi Partial Differential Equation
(HJPDE) reads
\begin{multline}\label{86}
H'_{0}=p_0+\frac{1}{4}\>F^{ij}F_{ij}-\frac{1}{2}\>\pi_{i}\pi^{i}+\pi^i\partial_iA_0+
\overline{\psi}\gamma^{\mu}eA_{\mu}\psi-\overline{\psi}(i\gamma^i\partial_i-m)\psi,
\end{multline}
\begin{equation}\label{87}
H'_1=\pi^{0}+H_1=\pi_0=0,
\end{equation}
\begin{equation}\label{88}
H'_{\psi}=p_{\psi}+H_{\psi}=p_{\psi}-i\overline{\psi}\,\gamma^0=0,
\end{equation}
\begin{equation}\label{89}
H'_{\overline{\psi}}=p_{\overline{\psi}}+H_{\overline{\psi}}=p_{\overline{\psi}}=0.
\end{equation}
Therefore, the total differential equations for the characteristic
(10), (11) and (12), obtained as
$$ dA^{i}=\frac{\partial
H'_{0}}{\partial\pi_{i}}\>dt+\frac{\partial
H'_{1}}{\partial\pi_{i}}\>dA^{0}+\frac{\partial
H'_{\psi}}{\partial\pi_{i}}\>d\psi+\frac{\partial
H'_{\overline{\psi}}}{\partial\pi_{i}}\>d\overline{\psi},$$
\begin{equation}\label{90}
~ =-(\pi^{i}+\partial_{i}A_{0})\,dt,\quad \quad \quad \quad \quad
\quad \quad \quad \quad \,\,\,\,\
\end{equation}

$$ dA^{0}=\frac{\partial
H'_{0}}{\partial\pi_{0}}\>dt+\frac{\partial
H'_{1}}{\partial\pi_{0}}\>dA^{0}+\frac{\partial
H'_{\psi}}{\partial\pi_{0}}\>d\psi+\frac{\partial
H'_{\overline{\psi}}}{\partial\pi_{0}}\>d\overline{\psi},$$
\begin{equation}\label{91}
~ =dA^0,\quad \quad \quad \quad \quad \quad \quad \quad \quad
\,\,\,\,\qquad \qquad \quad
\end{equation}

$$ d\pi^{i}=-\frac{\partial
H'_{0}}{\partial A_{i}}\>dt-\frac{\partial H'_{1}}{\partial
A_{i}}\>dA^{0}-\frac{\partial H'_{\psi}}{\partial
A_{i}}\>d\psi-\frac{\partial H'_{\overline{\psi}}}{\partial
A_{i}}\>d\overline{\psi},$$
\begin{equation}\label{92}
~ ~=(\partial_{l}F^{li}-e\overline{\psi}\gamma^i\psi)\,dt,\quad
\quad \quad \quad \quad \quad \quad \quad \quad \,\,\,\,\
\end{equation}

$$ d\pi^{0}=-\frac{\partial
H'_{0}}{\partial A_{0}}\>dt-\frac{\partial H'_{1}}{\partial
A_{0}}\>dA^{0}-\frac{\partial H'_{\psi}}{\partial
A_{0}}\>d\psi-\frac{\partial H'_{\overline{\psi}}}{\partial
A_{0}}\>d\overline{\psi},$$
\begin{equation}\label{93}
~ ~=(\partial_{i}\pi^{i}-e\overline{\psi}\gamma^0\psi)\,dt,\quad
\quad \quad \quad \quad \quad \quad \quad \quad \,\,\,\,\
\end{equation}

$$ dp_{\psi}=-\frac{\partial
H'_{0}}{\partial\psi}\>dt-\frac{\partial H'_{1}}{\partial
\psi}\>dA^{0}+\frac{\partial H'_{\psi}}{\partial
\psi}\>d\psi+\frac{\partial H'_{\overline{\psi}}}{\partial
\psi}\>d\overline{\psi},$$
\begin{equation}\label{94}
~ ~~~=-(i\gamma^i\partial_{i}+e\gamma^\mu
A_\mu+m)\overline{\psi}\,dt,\quad \quad \quad \quad \quad \quad
\quad
\end{equation}
and
$$ dp_{\overline{\psi}}=-\frac{\partial
H'_{0}}{\partial{\overline\psi}}\>dt-\frac{\partial
H'_{1}}{\partial{\overline\psi}}\>dA^{0}+\frac{\partial
H'_{\psi}}{\partial{\overline\psi}}\>d\psi+\frac{\partial
H'_{\overline{\psi}}}{\partial{\overline\psi}}\>d\overline{\psi},$$
\begin{equation}\label{95}
~=(-i\gamma^i\partial_{i}+e\gamma^\mu
A_\mu+m)\psi\,dt-i\gamma^0d\psi.\quad \quad
\end{equation}
\indent The integration condition $(dH'_\alpha=0)$ imply that the
variation of the constraints $H'_1,H'_\psi$ and
$H'_{\overline\psi}$ should be identically zero
\begin{equation}\label{96}
dH'_{1}=d\pi_0=0,
\end{equation}
\begin{equation}\label{97}
dH'_{\psi}=dp_\psi-i\gamma^0d{\overline\psi}=0,
\end{equation}
\begin{equation}\label{98}
dH'_{\overline\psi}=dp_{\overline\psi}=0,
\end{equation}
when we substituting  from Eqs. (94) and (95) into Eqs.(97) and
(98) respectively, we obtained as
\begin{equation}\label{99}
dH'_{\psi}=0,
\end{equation}
and
\begin{equation}\label{100}
dH'_{\overline\psi}=0,
\end{equation}
if and only if
\begin{equation}\label{101}
i\overline{\psi}\gamma^\mu(\overleftarrow{\partial}_\mu-ieA_\mu)+m\overline{\psi}=0,
\end{equation}
and
\begin{equation}\label{102}
i(\partial_\mu+ieA_\mu)\gamma^\mu\psi-m\psi=0,
\end{equation}
are satisfied. Then the set of equations (90, 92, 93) are integrable
and are just ordinary differential
equations and are set in the form
\begin{equation}\label{103}
{\dot A}^i=-\pi^{i}-\partial_{i}A_{0},
\end{equation}
\begin{equation}\label{104}
{\dot \pi}^i=\partial_{l}F^{li}-e\overline{\psi}\gamma^i\psi,
\end{equation}
\begin{equation}\label{105}
{\dot \pi}^0=\partial_{i}\pi^{i}-e\overline{\psi}\gamma^0\psi.
\end{equation}
These are the equations of motions with full gauge freedom. It can
be seen, from Eq. (91), that $A^0$ is an arbitrary (gauge
dependent) variable since its time derivative is arbitrary.
Besides that, Eq. (103) shows the gauge dependence of $A^i$ and,
Taking the curl of its vector form, leads to the known Maxwell
equation
\begin{equation}\label{106}
\frac{\partial\overrightarrow{A}}{\partial
t}=-\overrightarrow{E}-\overrightarrow{\bigtriangledown}(A_0-\alpha)\Rightarrow\frac{\partial\overrightarrow{B}}{\partial
t}=-\overrightarrow{\bigtriangledown}\times\overrightarrow{E}.
\end{equation}
\indent Writing $j^\mu=e\overline{\psi}\gamma^{\mu}\psi$ we get,
from Eq. (104), the inhomogeneous Maxwell equation
\begin{equation}\label{48}
\frac{\partial\overrightarrow{E}}{\partial
t}=\overrightarrow{\bigtriangledown}\times\overrightarrow{B}-\overrightarrow{j},
\end{equation}
while the other inhomogeneous equation
\begin{equation}\label{49}
\overrightarrow{\bigtriangledown}\cdot\overrightarrow{E}=j^0,
\end{equation}
follows from Eq. (105). Expressions (101) and (102) are the known
equations for the spinor $\psi$ and $\overline{\psi}$.\\
\indent Eqs. (12) and (86-89) lead us to the canonical action
integral as
\begin{multline}\label{109}
Z=\int
d^{4}x\bigg(-\frac{1}{4}\>F^{ij}F_{ij}+\frac{1}{2}\>\pi_{i}\pi^{i}+\pi^i{\dot
A}_i+\pi^i\partial_iA_0+i\overline{\psi}\gamma^{\mu}(\partial_\mu+ieA_{\mu})\psi-m\overline{\psi}\psi\bigg).
\end{multline}
Making use of equations (14) and (109), we obtained the path
integral as
\begin{multline}\label{110}
\left<out|S|In\right> = \int \prod_{i}\,dA^{i}\, d\pi^{i}\,
d\psi\, d\overline{\psi}\>exp\bigg[i\bigg\{\int d^{4}x
\bigg(-\frac{1}{4}\>F^{ij}F_{ij}+\frac{1}{2}\>\pi_{i}\pi^{i}\\+\pi^i{\dot
A}_i+\pi^i\partial_iA_0+i\overline{\psi}\gamma^{\mu}(\partial_\mu+ieA_{\mu})\psi-m\overline{\psi}\psi\bigg)\bigg\}\bigg].
\end{multline}
Integration over $\pi_i$ gives
\begin{multline}\label{111}
\left<out|S|In\right> =N \int \prod_{i}\,dA^{i}\, d\psi\,
d\overline{\psi}\>exp\bigg[i\bigg\{\int d^{4}x
\bigg(\frac{1}{2}({\dot
A}^i+\partial_iA_0)^2\\-\frac{1}{4}\>F^{ij}F_{ij}
+i\overline{\psi}\gamma^{\mu}(\partial_\mu+ieA_{\mu})\psi
-m\overline{\psi}\psi\bigg)\bigg\}\bigg].
\end{multline}

\section {Conclusion}
In this paper three different constrained fields systems are studied by using Hamilton-Jacobi formulation.
Firstly, he scalar field coupled to two flavours of fermions through
Yukawa couplings is discussed as constrained system by using Hamilton-Jacobi methods. The equations of motion are obtained without introducing Lagrange multipliers to the canonical
Hamiltonian.\\
\indent Then, we obtained the path integral quantization of the scalar field coupled
minimally to the vector potential by using the
canonical path integral formulation. The integrability
conditions $dH'_{0}$ and $dH'$ are satisfied, the system is
integrable, hence the path integral is obtained directly as an
integration over the canonical phase space coordinates$A^{i},
\pi^{i}, \varphi, P_{\varphi}, \varphi^{\ast}$ and
$p_{\varphi^{\ast}}$ without using any gauge fixing conditions.\\
\indent Finally, the path integral quantization of the electromagnetic field coupled to
a spinor is also obtained by using the canonical path integral
formalism. The integrability conditions $dH'_1$, $dH'_\psi$
and $dH'_{\overline{\psi}}$ are identically satisfied, and the
system is integrable. Hence, the canonical phase space coordinates
$(A^i, \pi^i), (\psi, p_\psi)$ and $(\overline{\psi},
p_{\overline{\psi}})$ are obtained in terms of the parameter
$\tau$. The path integral is obtained as an integration over the
canonical phase-space coordinates $(A^i, \pi^i)$ and $(\psi,
\overline{\psi})$ without using any gauge fixing condition. From
the equations of motion for this system, we obtained the
inhomogeneous Maxwell equation.\\
One can see many advantages of this path integral formalism, which are no
need to enlarge the initial phase-space by introducing unphysical
auxiliary field, no need to distinguish between first and
second-class constraints, no need to introduce Lagrange multiplierts, no
need to use delta functions in the measure as well as to use gauge
fixing conditions; all that needed is the set of Hamilton-jacobi
partial differential equations and the equations of motions. If
the system is integrable, then one can construct the reduced
canonical phase-space.\\

\end{document}